\newcounter{myctr}

\documentclass{ws-acs}
\usepackage{url}
\usepackage{graphicx}
\usepackage{float}
\begin{document}

\makeatletter
\def\@biblabel#1{[#1]}
\makeatother

\markboth{Huet S., Mathias J.D.}{Few self-involved agents among BC agents can lead to polarized local or global consensus}

%
\catchline{}{}{}{}{}
%

\title{FEW SELF-INVOLVED AGENTS AMONG BOUNDED CONFIDENCE AGENTS CAN CHANGE NORMS}

\author{\footnotesize Sylvie HUET\footnote{Typeset names in
10~pt Times Roman, uppercase. Use the footnote to indicate
the present or permanent address of the author.}}

\address{Irstea, Laboratory of Engineering for Complex Systems LISC, Address\\ 63172 Aubi\`ere, France\\
sylvie.huet@irstea.fr}

\author{Jean-Denis MATHIAS}

\address{Irstea, Laboratory of Engineering for Complex Systems LISC, Address\\ 63172 Aubi\`ere, France\\
jean-denis.mathias@irstea.fr}

\maketitle

\begin{history}
\received{(received date)}
\revised{(revised date)}
\end{history}

\begin{abstract}
Social issues are generally discussed by highly-involved and less-involved people to build social norms defining what has to be thought and done about them. As self-involved agents share different attitude dynamics to other agents \cite{WoodPool1996}, we study the emergence and evolution of norms through an individual-based model involving these two types of agents. The dynamics of self-involved agents is drawn from \cite{HuetDeffuant2010}, and the dynamics of others, from \cite{Deffuant2001}. The attitude of an agent is represented as a segment on a continuous attitudinal space. Two agents are close if their attitude segments share sufficient overlap. Our agents discuss two different issues, one of which, called main issue, is more important for the self-involved agents than the other, called secondary issue. Self-involved agents are attracted on both issues if they are close on main issue, but shift away from their peer’s opinion if they are only close on secondary issue. Differently, non-self-involved agents are attracted by other agents when they are close on both the main and secondary issues. We observe the emergence of various types of extreme minor clusters. In one or different groups of attitudes, they can lead to an already-built moderate norm or a norm polarized on secondary and/or main issues. They can also push disagreeing agents gathered in different groups to a global moderate consensus.
\end{abstract}


\section{Introduction}

In this paper, we define a norm as simply the behavior or opinion adopted by the majority of a group of people. We are not interested in descriptive norms but more in the injunctive norm emerging from our interactions with others and from the shared normative beliefs supporting it \cite{Bicchieri2014}. Whereas descriptive norms refer to what most people do, injunctive norms depict what most people approve of doing \cite{Farrow2017}, what is acceptable in a group. As a rule, the norm exerts social pressure to conformity in the group, and agents tend to avoid straying too far from it. However, norms do change, and there may also be a boomerang effect \cite{SchultzNolanCialdiniGolstein2007}. For instance, in most European countries, majority opinion on society issues like abortion, homosexuality and smoking has changed dramatically over the last 50 years. In agriculture, the norm has long been towards strong intensification, meaning most farmers attach high value to intensive practices and their high production outcomes regardless of whether or not they conform in terms of behavior. Today, however, more and more people and a minority of farmers disapprove of these practices, and this groundswell may bring about strong social change in our conceptions of agriculture. The research reported here sets out to understanding how such minorities emerge and how such changes can take place.

The scholarship has singled out the involvement of certain agents as a condition favoring resistance to the pressure to conformity. Muzafer Sherif especially outlined the role of ego-involvement in persuasion: ``... Regardless of the discrepancy of the position presented, we predict that the more the person is involved in the issue (the more important it is to him), the less susceptible he will be to short-term attempts to change his attitude" \cite{Sherif1968}. Ego-involvement ``refers more generally to the involvement of the self or personal involvement. Ego-involving topics are those that have intrinsic importance and personal meaning ... Important or involving issues were those that had self-relevance" \cite{PettyCacioppo1992}.

In his theories of social comparison \cite{Festinger1954} and cognitive dissonance \cite{Festinger1957}, Leon Festinger highlighted the need to be close, to be a member of a group, and to be cognitively consistent. Here, building on these seminal works, we propose an agent-based model inspired by \cite{WoodPool1996} and social identity theories \cite{Brewer} \cite{TajfelTurner1979} \cite{Turner1984} to explore how the strong ego-involvement of a part of the population may be responsible for the evolution of a norm. The model agents can be involved or not in an issue being discussed, and the corresponding dynamics are based on \cite{Deffuant2001} and \cite{HuetDeffuant2010}. Two issues are discussed. Some agents are involved in neither of them, while others are involved in one issue, called main issue, but not in the other, called secondary issue.

The simulations show the emergence of small sets of agents sharing common extreme opinions on the issues and differing from the majority. Under some conditions, these agent clusters can lead a moderate majority to follow their extreme opinion on the main or the secondary issues. This drift to the extreme can also take place when the moderate majority is split in several different groups of opinions, each having their own norm. However, we observe that the polarization on the secondary issue can be very slow. In all these cases, we observe that a moderate norm changes for a more extreme one. In what follows, we start by giving an overview of the model, then go on to describe more details of the results and explain how they emerge, and finish by summarizing the main results and discussing the interpretations .


\section{State of the art}

Global overviews of the social psychology literature and agent-based models related to social influence can be found in \cite{Castellano2009} \cite{Mason2007} \cite{Flache2017review}. The vast majority of existing models fail to consider the involvement of the agent in the issue discussed, despite it long being recognized as an important variable by the social psychology scholarship \cite{Allport1943} \cite{SherifHovland1961} \cite{Sherif1968} \cite{Thomsen1995}.

A few models do consider agent involvement to some extent. The bounded confidence model \cite{Deffuant2001} \cite{HegselmannKrause2002} is quite consistent with social judgment theory, which states that positive influence (i.e. opinion or attitude attraction) is a function of the discrepancy of opinions between two agents, bounded by the influencee’s level of ego-involvement in the discussed issue \cite{SherifHovland1961}. The ``confidence segment" may thus be related to the involvement in an issue - the larger the segment, the less involved the agent.

Assuming that agents can sometimes shift away from each other as shown by \cite{Whittaker1963} or \cite{SherifHovland1961}, \cite{Mark2003} \cite{JagerAmblard2004} \cite{JagerAmblard2005} and \cite{FlacheMacy2011} have proposed models with one threshold for attraction and the same or another threshold for rejection. Overall, they have modeled attitude shifts as proportional to the distance to the threshold relevant for rejection. These thresholds can be also interpreted as defining the level of the agent’s ego-involvement. Another investigation by simulation showing potential links between involvement and attitude dynamics can be found in \cite{Salzarulo2006} with a model derived from self-categorization theory \cite{TurnerHogg1987}. More on more-or-less similar models can be found in a recent review  \cite{Flache2017review} that concluded on a need to compare existing models. In line with this recommendation, below we give an in-depth description of models potentially relatable to ego-involvement and sufficiently related to be meaningfully compared.

Bounded confidence models can show convergence to extreme or moderate consensus. The simulation typically starts with a population including two types of agents: extremists who have extreme and very certain attitudes (they are never influenced by others), and moderates whose fairly uncertain attitudes are uniformly distributed and who can easily be influenced. Depending on the initial number of extremists and the initial uncertainty of the moderates, three types of convergence emerge: central clusters, double extreme clusters in which almost everyone has adopted one or the other extremist attitude, and a single extreme cluster implying that everyone shares the same extremist attitude \cite{Deffuant2002} \cite{Deffuant2006} \cite{DeffuantWeisbuch2007}. A single extreme cluster emerges from an initial convergence to a central opinion of moderate agents which is then pulled by the extremist agents. This  can be related to a norm change. Recently, \cite{Mathias2016,Mathias2017} show that a new stationary state appears when the uncertainties do not change over time and when the moderates have very large uncertainties, i.e. the attitudes of moderate agents keep fluctuating without clustering, but the distribution of the attitudes remains stable over time. These models assume that attraction to the extreme minority is the main driver of norm change. Importantly, they suppose that this extreme minority is initially present in the population - they do not explain how these initial extremists appear.

These models are in line with social psychology scholarship that assumes attraction to an extreme minority attitude \cite{MoscoviciZavalloni1969} or an extreme minority group \cite{EaglyChaiken1993} \cite{WoodOuellette1994} \cite{CranoPrislin2006} is the main driver of polarization. This is also tied to ego-involvement, since social judgment theory \cite{SherifHovland1961} from which the bounded confidence models are largely inspired, assumes that level of ego-involvement influences the sensitivity of the agents to persuasion. It argues for an influence that increases and then decreases with the discrepancy between the attitude of influencer and the attitude of the influence. The level of ego-involvement defines the inflexion point of the influence. Greater ego-involvement leads to thicker discrepancy before the inflexion point, and thus lower susceptibility to opinion change.

Another hypothesis to explain norm change can be found in a willingness of agents to differentiate from others or at least some others. This is particularly salient when agents are in deep disagreement on an important issue, typically one that involves their identity, in which case they tend to be willing to differentiate on other, less important issues. Indeed, several studies and theories suggest that polarization is connected with identity issues \cite{Brewer} \cite{TajfelTurner1979} \cite{Turner1984}. The set of experiments presented in \cite{WoodPool1996} suggests that level of self-relevance of the source-group for the agent is a key factor in the rejection mechanism. Indeed, the experiments show that in cases of agreement on a minor issue and disagreement on a highly self-relevant issue (i.e. group membership in the experimentation), a subject tends to shift away from alters  position on the minor issue. Moreover, in cases of agreement on a highly self-relevant issue, the subjects tend to be attracted on minor issues whatever the level of preliminary disagreement. \cite{HuetDeffuant2010} proposed an agent-based model inspired by \cite{WoodPool1996} considering a dynamics for highly self-involved agents (considering the degree ego-involvement in an issue as very similar in concept to the self-relevance of an issue). The authors consider peer  interactions in which two agents influence each other on attitudinal dimensions, assuming that one attitude is important (main) and the other is secondary. On each of the two dimensions (i.e. two attitudes), they suppose that the attitudes can take continuous values between -1 and +1. When the agents are close to each other on the main dimension, they tend to attract each other on the secondary dimension, even if their disagreement is strong. When they disagree on the main dimension and are close on the secondary, then they tend to reject each other on the secondary dimension. The model includes two thresholds: the $u_m$ threshold on the main dimension, and the $u_s$ threshold on the secondary dimension.

This model leads to approximately $1/u_m$ main opinion clusters, while in the classical bounded confidence model, this number is approximately $1/u_m*1/u_s$. The reason for this significant difference is that in the model proposed in \cite{HuetDeffuant2010} the clustering is only driven by the main dimension with its threshold $u_m$; the number of secondary dimensions does not influence the number of clusters \cite{HuetDeffuant2010}. This makes the number of clusters independent from the number of secondary dimensions, and simplifies the perception of the social space (the smaller number of clusters).

Simulations with this model also show the existence of minor clusters containing a very small number of agents, when the parameter ruling the speed of attraction is sufficiently large (as is the case in the following study). These minor clusters have already been observed in the bounded confidence model. They are located between the main clusters and at the extremes of the opinion space, especially for large values of confidence threshold $u$ leading to one major cluster containing at least 95\% of the population \cite{Laguna2004} \cite{Lorenz2007}. These minor clusters are important in the new model that we describe later in this paper.

\cite{HuetDeffuant2010} also observe a polarization on the secondary dimension due to rejection. This polarization often occurs after a first stage of clustering in a moderate opinion and then they  migrate to an extreme attitude. However, this dynamics takes place only when all agents share the main dimension as highly self-relevant, as in the experiments of  \cite{WoodPool1996}. Of course, in practice it is unlikely to find a situation where all the agents of a population define themselves through the same salient social issue. 

\cite{WoodPool1996} (pp. 1190, 1191) adds that if the participant is not highly negatively self-involved in the source-group of the source, he will not change his opinion when he agrees with the source on the secondary issue but not on the main issue. Instead, he decreases the relevance of the source (i.e. the level of importance assigned to the sources group) in a way to not differentiate from it in the future. They add that participants who were relatively indifferent to the group source demonstrated little shift in their attitude judgments (p. 1186). Here, then, and for the sake of simplicity, we assume that some agents are indifferent to their group membership and do not change their opinion if they disagree on only one dimension and that self-relevance, or level of importance given to the main attitude, is constant.

In the following, we consider a mixed population composed of a static set of highly-self-involved agents (called HSI agents) in one main attitude and the complementary set of no-more-important-attitude agents (called non-HSI agents). The dynamics of the HSI type of agents is based on \cite{HuetDeffuant2010} while the dynamics of the non-HSI type is based on a classical 2D bounded confidence model \cite{Deffuant2001}. Our populations of agents have attitudes initially drawn uniformly. We simulate their evolution, for different values of the parameters, in order to compare the behavior of a mixed population to the behavior of pure population of highly involved agents or non-highly-involved  agents. Analysis of the simulation found the emergence of several types of extreme minor clusters. They can lead a population that has converged to a moderate norm in a first stage to an extreme norm on both the minor and fundamental ones. This phenomenon can drive the majority of the population or only some groups of opinions. We also observe that several groups polarized on secondary issues ultimately, after a very long time, adopt a unique moderate norm. Some of these observations are specific to mixed populations. Moreover, a small number of self-involved agents in the population is enough to significantly reduce the number of clusters. 

In what follows, we start by giving an overview of the model, then go on to describe more details of the results and explain how they emerge, and finish by summarizing the main results and discussing the interpretations.


\section{The model and experimental design}

We consider a population of $N$ individuals, part of them considered highly-self-involved in the main dimension while the other part is not. The model includes four parameters: $h$, the proportion of highly-self-involved people that share very similar dynamics to the people considering group membership as highly self-relevant in \cite{Whittaker1963}, $u_m$ and $u_s$,  thresholds for opinion change on the main and the secondary dimension, respectively, and $\mu$ ruling the intensity of influence at each meeting (comprised between 0 and 0.5). An individual has two attitudes, $x_m$ (on the main dimension) and $x_s$ (on the secondary dimension), taking real values between -1 and +1. During an iteration, a pair of individuals \textit{X} and \textit{Y} is randomly chosen and can influence each other. The algorithm is the following:

\begin{itemlist}
\item Choose a random couple (\textit{X,Y}) of individuals in the population; 
\item \textit{X} and \textit{Y} change their attitudes at the same time, according to the influence function corresponding to their status, i.e. the potential influencee is a non-highly-self-involved (non-HSI) agent or a highly-self-involved (HSI) agent. 
\end{itemlist}

\subsection{Influence on a non-highly self-involved agent (non-HSI)}

The calculation of the influence of \textit{Y} on \textit{X} if \textit{X} is a non-HSI agent (of course the influence of \textit{X} on \textit{Y} if \textit{Y} is a non-HSI agent is found by inverting \textit{X} and \textit{Y}) is as follows. Let ($x_m$, $x_s$) and ($y_m$, $y_s$) be the attitudes of \textit{X} and \textit{Y}, respectively. The influence of \textit{Y} on \textit{X} is not null if $\left|x_m-y_m\right| \leq u_m$ and $\left|x_s-y_s\right| \leq u_s$ indicating that agent \textit{X} agrees with \textit{Y}. Both attitudes of \textit{X} are going to get closer to those of \textit{Y}, proportionally to the attitudinal distance on each dimension:
\begin{equation}
x_m(t+1)= x_m(t)+\mu(y_m(t)-x_m(t))
\end{equation}
\begin{equation}
x_s(t+1)= x_s(t)+\mu(y_s(t)-x_s(t))
\end{equation}

For every other case of partial agreement (only close on one dimension) or total disagreement (far on two dimensions), \textit{X} remains indifferent to \textit{Y} and does not change opinion. 

\subsection{Influence on a highly self-involved agent (HSI)}

The calculation of the influence of \textit{Y} on \textit{X} if \textit{X} is a HSI agent (of course the influence of \textit{X} on \textit{Y} if \textit{Y} is a HSI is found by inverting \textit{X} and \textit{Y}) is based on \cite{HuetDeffuant2010} as follows. Let ($x_m$, $x_s$) and ($y_m$, $y_s$) be the attitudes of \textit{X} and \textit{Y} respectively. We first consider the main attitude dimension.

If $\left|x_m-y_m\right|$ $\leq$ $u_m$, agent \textit{X} agrees with \textit{Y} on the main dimension. Both attitudes of \textit{X} are going to get closer to those of \textit{Y}, proportionally to the attitudinal distance on each dimension:
\begin{equation}
x_m(t+1)= x_m(t)+\mu(y_m(t)-x_m(t))
\end{equation}
\begin{equation}
x_s(t+1)= x_s(t)+\mu(y_s(t)-x_s(t))
\end{equation}

Indeed, whatever the agreement level on the secondary dimension is, \textit{X} is going to be globally closer to \textit{Y}. If it was already close, it gets closer. 

If $\left|x_m-y_m\right|$  $>$ $u_m$, then agent \textit{X} disagrees with \textit{Y} on the main dimension, and if  $\left|x_s-y_s\right|$ $\leq$ $u_s$, then agent \textit{X} feels it is too close to \textit{Y} on the secondary dimension, because of their disagreement on the main dimension. \textit{X} solves this conflicting situation by moving away from \textit{Y} on this dimension. The attitude change is proportional to the distance to reach the rejection threshold: 

\begin{center}
if $\left(x_s-y_s\right) < 0$, then $x_s\left(t+1\right)=x_s\left(t\right) - \mu\left(u_s-\left(y_s\left(t\right)-x_s\left(t\right)\right)\right)$
\end{center}
\begin{center}
	else if $\left(y_s\left(t\right) \neq x_s\left(t\right)\right)$, then $x_s\left(t+1\right)=x_s\left(t\right) + \mu\left(u_s+\left(y_s\left(t\right)-x_s\left(t\right)\right)\right)$ 
	\end{center}
	\begin{center}
		else $x_s\left(t+1\right) = x_s\left(t\right)+$ sign$\left(\right) \mu u_s$ 
\end{center}

where sign() is a function returning -1 or 1, each with a probability 0.5. This sets the direction of the opinion change chosen at random when \textit{X} and \textit{Y} perfectly agree.

In the other cases, \textit{X} is not modified by \textit{Y}.
Moreover, we confine the main attitude in the interval [-1, +1]. The secondary attitude can be confined in the interval [-1, +1], or totally unbounded. These two cases allow, in an experimental design considering both, an understanding of the impact of the confinement.
The attitude of \textit{Y} is calculated in the same way, considering the situation of the meeting with \textit{X} and if \textit{Y} is HSI.

\subsection{Experimental design and the measured indicators}

We run the model, for different values of the parameters, in order to compare the behavior of a mixed population to the known behaviors of a pure population of HSI or non-HSI agents. We vary the following parameters considering 10 replicates for each set of values and 10,000 agents. Using 10 replications is sufficient for our study, since there is little variation in outcome measures between runs: the average standard deviation of our main indicator, i.e. the average absolute opinion of the population varying between 0 and 1, is 0.06. $u_m$ and $u_s$ vary from 0.05 to 1 in increments of 0.05. Most of the study is done using \textit{h} equal to 0.1. However, we also vary \textit{h} to show that the emerging patterns are sensitive to this parameter for what is observed on the secondary attitude dimension. For a same set of parameter values, we consider two cases for the purpose of comparison: (1) on the secondary dimension, the opinion is confined between [-1;+1]; (2) it is not confined, nor on the main, neither on the secondary dimension. However, this is only during the simulation, since in both cases the initial distribution is confined in [-1;+1].

While the basic study is done with 10,000 agents, the figures representing one population state in what follows can be built on a population of 25,000 or 5,000 agents (figures 1 to 4 and 8). This is only to make the figures easier to read and does not impact the qualitative results observed. The figures giving an overview of the behavior of the model (figures 5, 7, 9 and 10) are built from a population of 10,000 agents.

We arbitrarily stop the exploration of $u_m$ and $u_s$ at 1 for this first study. Similarly, $\mu$, the parameter ruling the intensity of the influence, remains constant on the two dimensions for every experimental condition, at a value of 0.5.

The second case was studied to prove that the bounds are not responsible for the polarization. Indeed, due to bounds, some clusters of agents unable to adopt an attitude far enough from others might explain why some other clusters polarize and are finally located further than expected. Two indicators are measured during the simulation and then averaged over the replicates in order to define the number of norms we observe as well as how moderate they are. These indicators are:

\begin{itemlist}
\item the number of clusters (which is classically counted by considering an agent as a member of a cluster if it is located at a distance smaller than a threshold from at least one member of the group considered); each cluster has its own norm, which is the average opinion of its members; 
\item the average absolute opinion on each dimension so as to define how polarized the population is. Polarization is classically defined as an extremization of attitudes after some discussions in a group compared to the attitudes initially held by the in-group people. It is classically measured by the average opinion before and after a discussion. When the average opinion has increased in absolute value, the discussion is assumed to have led to polarization, in a process that is both individual and collective. 
\end{itemlist}

\textbf{The interpretation of the level of moderation of the norm(s) and its evolution from the average absolute opinion depends on the number of clusters}. We know this model is ruled in terms of group number by the dynamics of the bounded confidence model. From this model, we are able to define the expected absolute average opinion of the population when it does not polarize. We thus know the evolution has to be interpreted differently when the final number of groups is one and when it is at least two or more: 
\begin{itemlist}
\item when all the agents converge to a unique group, we know that they tend to a moderate central opinion which is 0.5 in our model. Since the agents’ opinions are initialized at random, the average absolute opinion is valued 0 at the beginning, as the opinion case goes from -1 to +1. Then, for a unique opinion group (a unique norm), a final value of the average absolute opinion higher than 0 indicates a polarization of the norm. If the average absolute opinion is higher than 0.5, this means the population gathered into one moderate group (with an average absolute opinion close to 0) before adopting an extreme norm (this behavior will be illustrated in the following);
\item when agents are gathered in several groups at the end of the simulation, the initial average absolute opinion is still valued 0.5 but the expected moderate final average absolute opinion varies from 0.4 to 0.5 depending on whether the number of clusters is odd or even, respectively. Then we can diagnose the level of polarization of the groups - and each of their corresponding norms - using these references.  
\end{itemlist}


\section{The trajectories to understand the global dynamics}

\begin{figure}[ht]
\includegraphics[width=6.0cm]{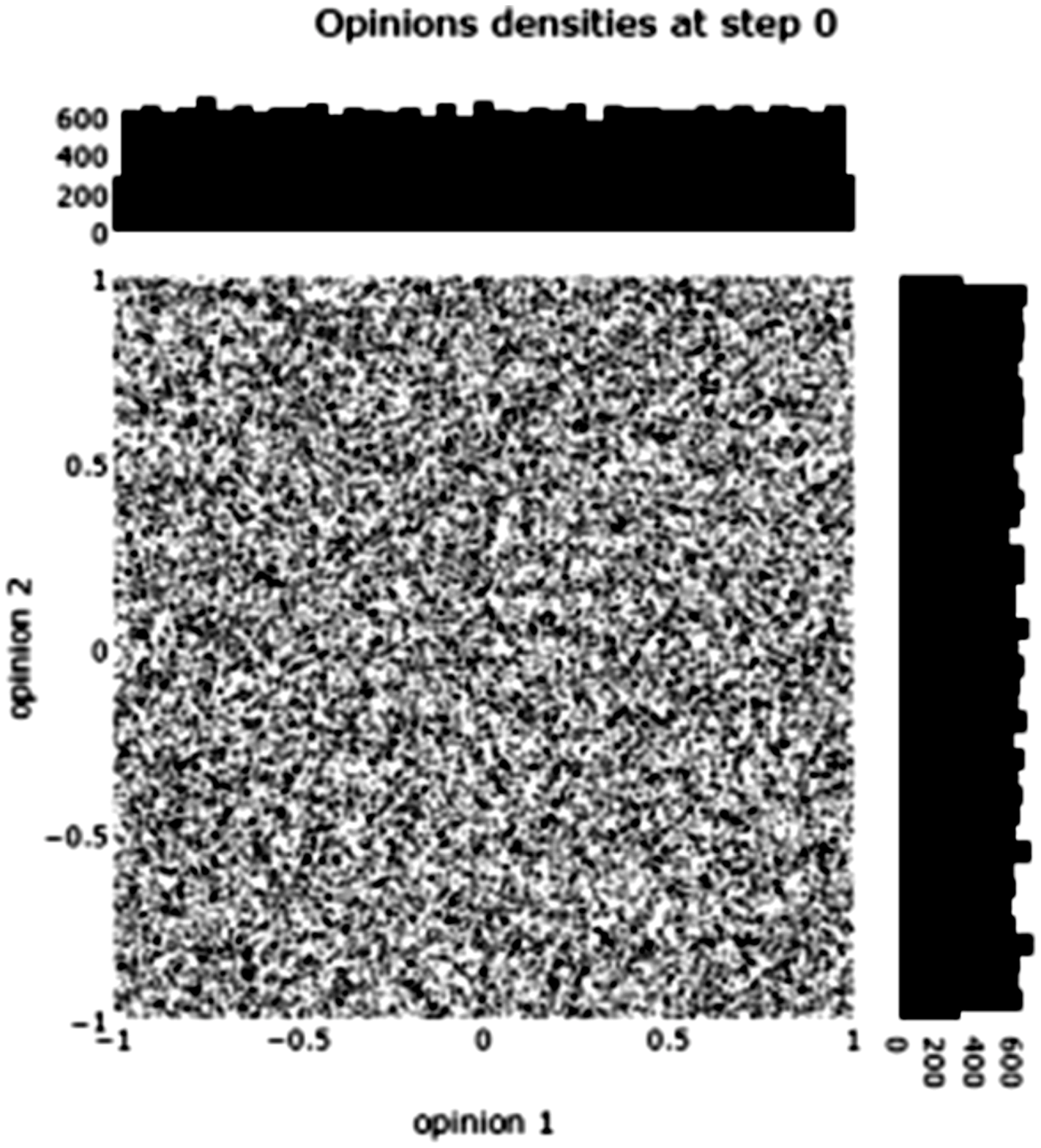}
\includegraphics[width=6.6cm]{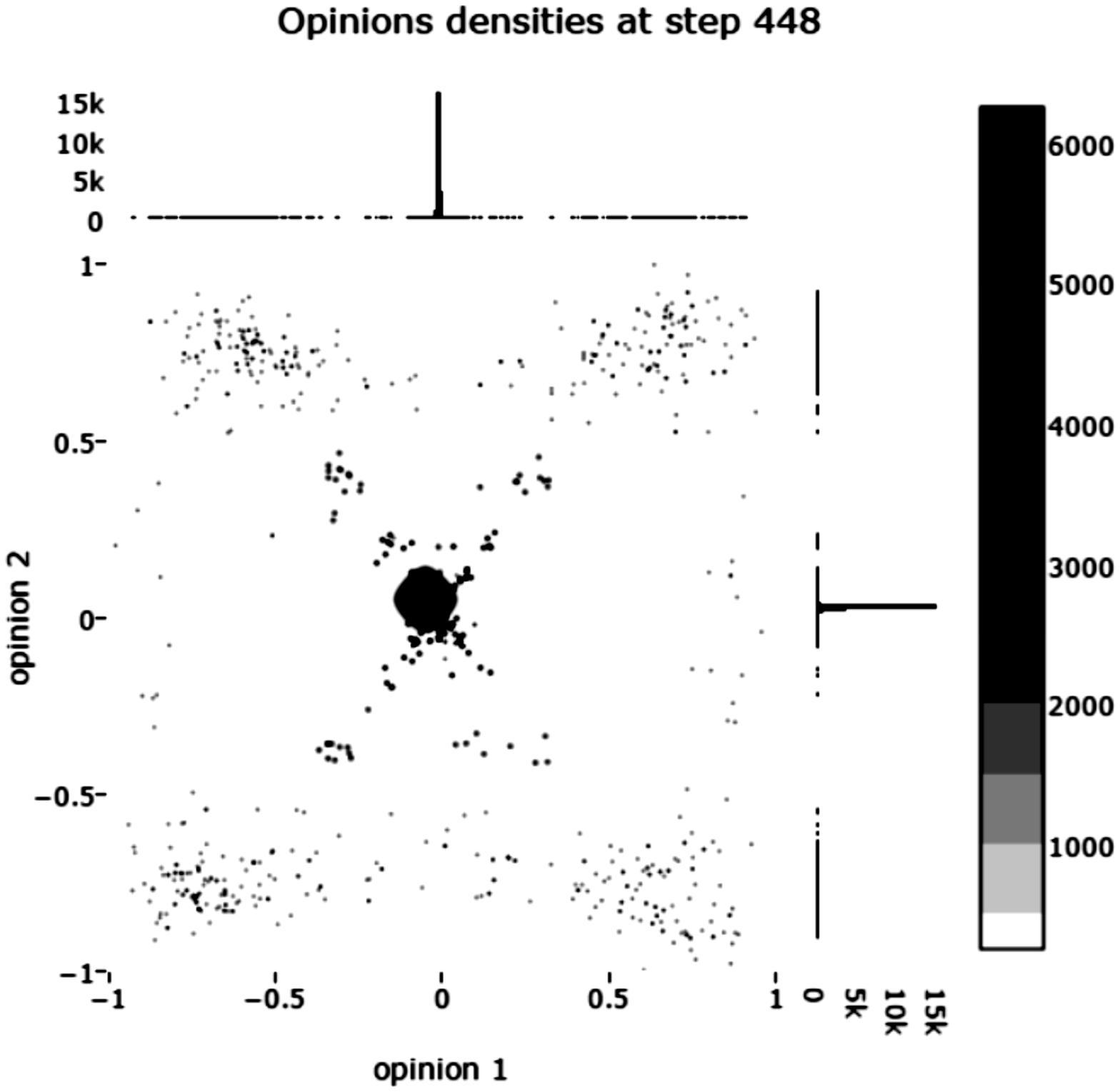}
\caption{Opinion states for 25,000 agents (25,000 agents were chosen for illustration purposes to make the dynamics more visible on the graph): on the left, initial random opinion; on the right, at time-step 448. Each dot represents one couple of opinion  ($x_m$,$x_s$) of an agent. On the x-axis, opinion 1 is the main opinion. On the y-axis, opinion 2 is the secondary dimension. The attitudes on the secondary dimension are confined in [-1, +1]. Small dots are non-HSI agents, and large dots are HSI agents. The density distributions of opinions on each dimension are visible on the right and at the top of the right-hand graph. The model can be run at http://motive.cemagref.fr/lisc/bc/ and thus the trajectories seen in color.}
\label{fig1}
\end{figure}

This section shows typical trajectories of the model corresponding to the emergence of a first norm which then goes on to change due to the dynamics. These trajectories illustrate the main principles of the model. They have been selected for illustration purposes considering the most frequent steady state of the bounded confidence model: the moderate consensus for which every or almost every agent adopts a moderate opinion. Figure \ref{fig1} presents the initial state as well as the type of graph used to show the trajectories of the model (on the left). All along the simulation, dots on graphs represent agents’ opinions: small dots represent the non-HSI agents, and large dots represent the HSI agents. Since the clusters are sufficiently formed, they are also represented in density over the parameter space (on the right). The scale for the density is on the right of the figure. The model can be run online at http://motive.cemagref.fr/lisc/bc/ and thus the trajectories seen in color (the code is also available). All the graphs in this section consider a population with 25,000 agents to make them more readable, and a confined secondary attitude.

\subsection{A moderate consensus on the main dimension polarizes due to a minority effect}

\begin{figure}[!htb]
\begin{center}
\includegraphics[height=15.6cm]{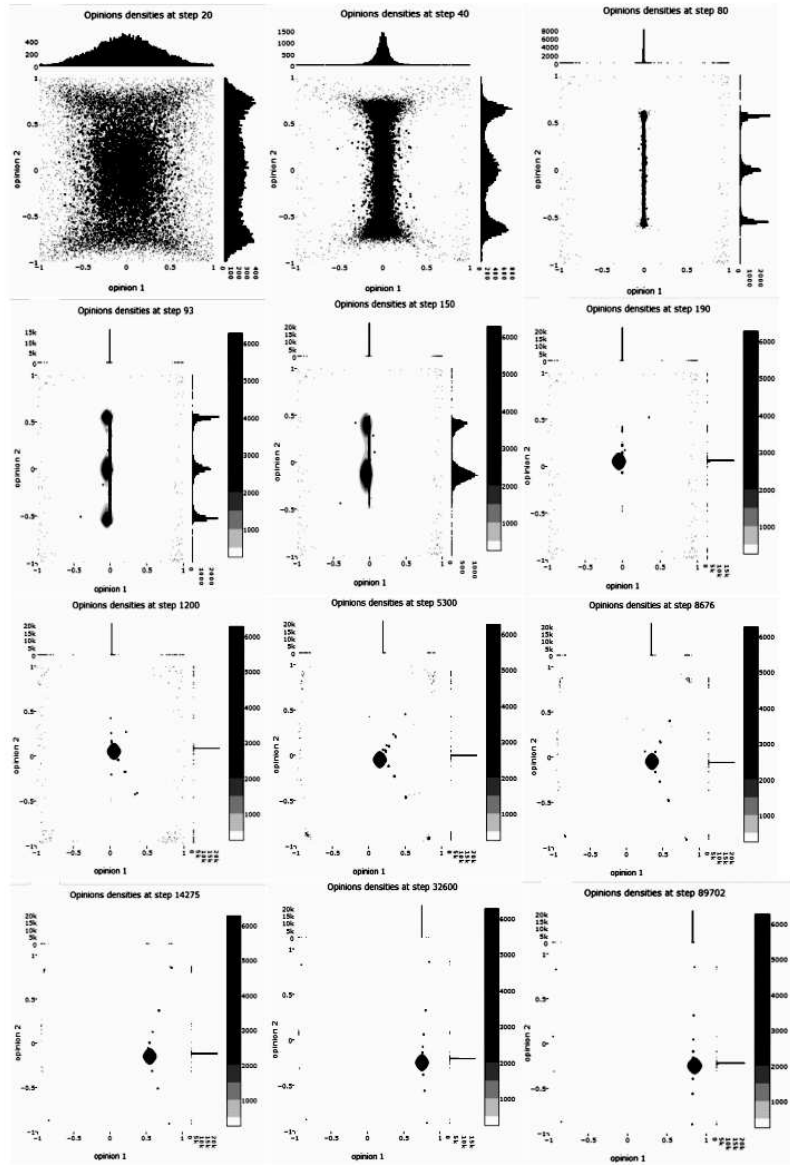}
\caption{From a moderate majority to an extreme majority on the main dimension. Stable state for 25000 agents, $u_m$ = 0.8, $u_s$ = 0.3, \textit{h} = 0.1 (large dots are HSI and small dots are non-HSI agents). From top left to bottom right, time: 20, 40, 80, 93, 150, 190, 1200, 5300, 8676, 14275, 32600 and 89720. On $x$-axis, opinion 1 is the main opinion. On $y$-axis, opinion 2 is the secondary dimension. The attitudes on the secondary dimension are confined in [-1, +1]. More details on how to read figures can be found in the caption to Fig. \ref{fig1}.}
\label{fig2}
\end{center}
\end{figure}

Figure \ref{fig2} presents the first unexpected impact of the presence of HIS agents in the population. After an expected convergence, since $u_m$ is higher than 0.5, to a moderate consensus among almost all the agents, this consensual group is seen to polarize due to the impact of extreme minorities. The trajectory is mainly ruled by attraction between agents. Due to the high value of $u_m$ which implies strong interagent influence, agents quickly gravitate to the center (from step 20 to step 190), and some agents are forgotten by the dynamics on the extreme values of the main dimension (see the histogram representation of the main dimension on the figures). HSI agents gravitate to center quicker than non-HSI agents; they only need to be close to the influencer on the main dimension to converge to the center. Non-HSI agents are slower due to the double constraint for influence (close on the two dimensions), and so they are then the ones forgotten by the dynamics on the extremes. They gather together and constitute the ``minor clusters"’ referred to in the state-of-the-art section. For $u_s$ $<$ $u_m$, the initial centralization on the secondary dimension is slower than the centralization on the main dimension (see figures from step 20 to 190). Then, there are less forgotten agents at the extremes on the secondary dimension. Moreover, HSI agents gravitate to center quicker than non-HSI agents (see larger dots in the center compared to small ones around at steps 20 to 80).

We observe a second period in the simulation from step 80 to step 190. This is a transitory in-group polarization on the secondary dimension. Almost all the agents are gathered in the central cluster defined by the main dimension. However, regarding the secondary dimension, non-HSI agents, which composed the majority of the central cluster, are gathered in three clusters (see the density distribution on the right of the figures, step 80 and 93). The HSI agents, which are members of this central cluster due to their dynamics, assess the non-HSI agents, which are also members of the central cluster, as being member of their unique cluster also on the secondary dimension. They are then influenced by them, and continuously move from one non-HSI secondary dimension cluster to another. At the same time, they influence in return the non-HSI agents, progressively attracting them to the center, reducing the number of clusters on the secondary dimension to two (see density distribution on the figure at step 150), and finally gathering them into a unique central cluster (step 190).

In a third period, since everyone is gathered in a unique large centered cluster, a first moderate norm has been built. However, this moderate cluster slowly oscillates in the opinion space due to HSI agents who want to be closer to forgotten agents on the borders since they consider them as members of their own group. This movement leads the larger cluster to a polarization that is only stopped when all forgotten people who are close on the main dimension are gathered in it (see from step 190 to the end). The major cluster, which had been moderate at time 190 and for a long time, has ultimately changed norm by the end to finish with a relatively extreme norm.

It is obvious that HSI agents represented by the larger dots are attracted on both dimensions by the extreme corners of non-HSI agents. To sum up, \textbf{it is the stability of the emerging small corner clusters which is responsible for the polarization of the larger cluster. This stability comes from non-HSI agents which require closeness on both the main and secondary dimensions in order to be influenced}. It allows the extreme corners of non-HSI agents to attract the HSI agents previously located in the center. \textbf{This first type of polarization is entirely due to the attraction process and does not depend on the rejection process}.

\begin{figure}[!htb]
\begin{center}
\includegraphics[width=11.3cm]{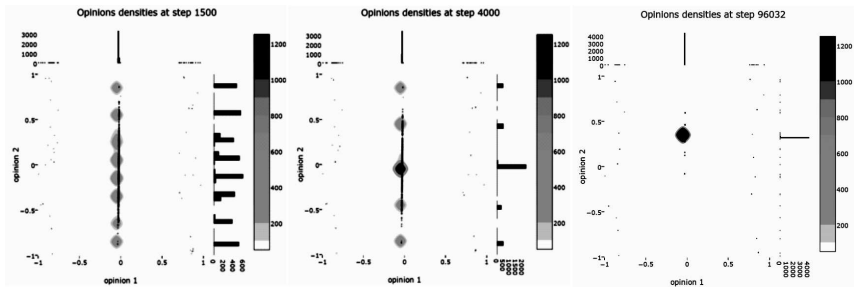}
\caption{Long transitory in-group polarization on the secondary dimension  $u_m$ = 0.7, $u_s$ = 0.1, \textit{h} = 0.1, 5000 agents. The clustering process on the secondary dimension is very long when $u_s$ is small. Then, the agents stay a long time within several clusters on the secondary dimension inside a unique cluster based on the main dimension (see time 1500 and 4000 on the left and the centered graph). However they ultimately form a unique cluster after a very long time (step 96032 on the right). On $x$-axis, opinion 1 is the main opinion. On $y$-axis, opinion 2 is the secondary dimension. The attitudes on the secondary dimension are confined in [-1, +1]. More details on how to read figures can be found in the caption to Fig. \ref{fig1}.}
\label{fig3}
\end{center}
\end{figure}

The transitory in-group polarization on the secondary dimension observed during the second period can last for long as shown in figure \ref{fig3}. It occurs when $u_s$ is small (and $u_m$ $>$0.5). It is due to the time required to reach the convergence, which increases when $u_s$ decreases. At smaller $u_s$ values, the unique central cluster covers almost all the secondary dimension for a long time, composed of various subclusters of non-HSI agents which are joined altogether by moving HSI agents (see Fig. \ref{fig5}). At the end, everyone is gathered in the center due to the impact of the HSI agents. This perfectly describes how HSI increases the level of cohesion of the population compared to a population of non-HSI agents (i.e. compared to the classical bounded confidence model). 

\subsection{A moderate consensus on the secondary dimension polarizes due to another minority effect}

\begin{figure}[ht]
\begin{center}
\includegraphics[width=11.3cm]{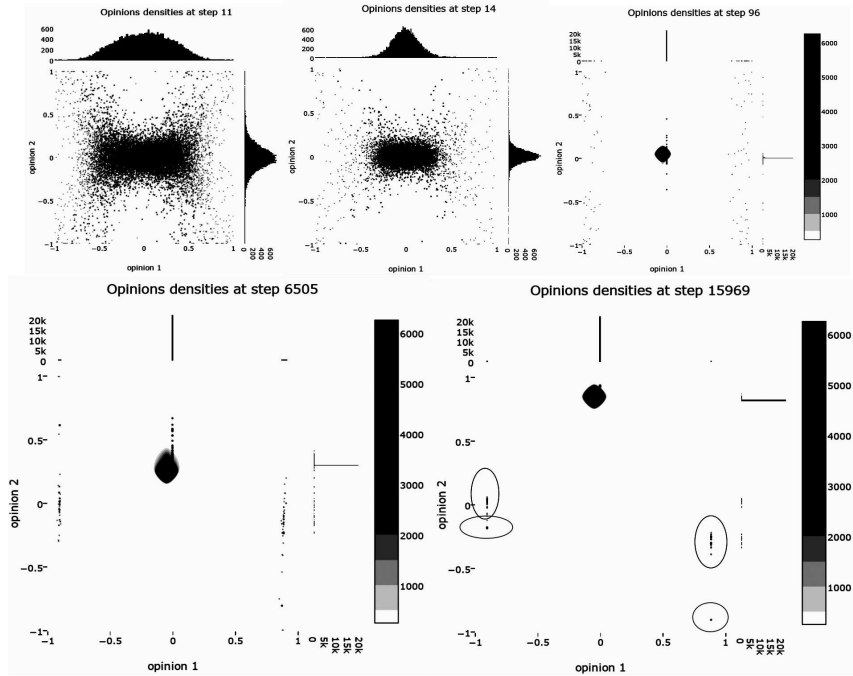}
\caption{From a moderate majority to an extreme majority on the secondary dimension. Stable state for 25000 agents, $u_m$ = 0.7, $u_s$ = 1.0, \textit{h} = 0.1 (large dots are HSI agents and small dots are non-HSI agents). Top circles indicate extreme cluster of non-HSI agents and bottom circles indicate HSI agents. From top left to bottom right, time: 11, 14, 96, 6505 and 15969. On the $x$-axis, opinion 1 is the main opinion. On the $y$-axis, opinion 2 is the secondary dimension. The attitudes on the secondary dimension are confined in [-1, +1]. More details on how to read figures can be found in the caption to Fig. \ref{fig1}.}
\label{fig4}
\end{center}
\end{figure}

Figure \ref{fig4} shows how a similar change can also occur on the secondary dimension in a mixed population. For our particular case, compared to the minority effect leading to norm change on the main dimension, agents in the central cluster gather very fast on the secondary dimension (see steps 11 and 14). As in the previous subsection, some non-HSI agents remain forgotten on the extremes, especially on the main dimension because $u_s$ is so large on the secondary dimension that the convergence to the center happens very quickly and does not forget anyone except if the closeness conditions is not respected on the main dimension. Moreover, agains since $u_s$ is very large, agents are very close to each other on the secondary dimension before being close on the main. In such situation, HSI agents located at one extreme of the large cluster on the main dimension shift away from agents (HSI or non-HSI) located at the other extreme of this same large cluster, since they do not consider them as members of the same group while feeling close on the secondary dimension (see steps 11 and 14, HSI agents represented by the larger dots). Some HSI agents polarize progressively to the extreme by shifting away from the other HSI agents and the forgotten non-HSI agents located close to the other extreme, while the other agents  have gathered to a major cluster with a moderate norm (see step 96). 

Then, these HSI and non-HSI agents gather together to create extreme minor clusters (see step 6505). From the point of view of the HSI agents, members of the largest initially central cluster, the members of the minor clusters are considered as close on the secondary dimension while far on the main dimension. The HSI agents of the largest cluster then shift away from them while simultaneously attracting the non-HSI agents of the largest cluster. On the other hand, extreme HSI agents of the minor clusters do the same (see bottom circles on fig. \ref{fig4}). This makes the large cluster polarized on the secondary dimension until the minor clusters are no longer considered close while minor clusters do the same. Then, once again, after having an initially moderate norm, the major cluster adopts an extreme norm, but on the secondary dimension. \textbf{This type of polarization depends on rejection (it does not appear for a similar model with no rejection)}.


\section{The global properties of the dynamics}

This section overviews the stable states reached by the model presented in section 3. We especially want to see which norm is reached and whether it changed during the simulation.  

\subsection{When does the norm change on the main dimension?}

In the following, we address our research question by looking on the main dimension and the secondary dimension in turn. We draw maps of the average absolute opinion at the end of the simulation for various $u_m$ and various $u_s$ on the main and on the secondary dimension. As explained in subsection 3.3, interpretation of the degree of gray depends on the number of clusters at the end of the simulation. Two areas should be distinguished: $u_m$ $>$ 0.5 giving more than one cluster on the left of the map; and $u_m$ $<$ 0.55 giving one major cluster on the right of the map. Perfect moderation of the final opinions is indicated in white on the right but mid-grey on the left. For both sides, all other gray degree show various levels or polarization.

\begin{figure}[ht]
\begin{center}
\includegraphics[width=10cm]{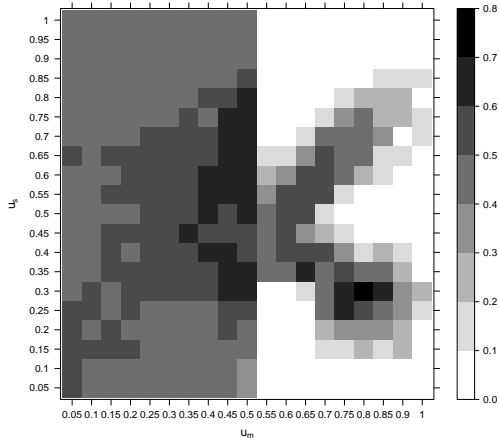}
\caption{Variation of average absolute opinion on the main dimension for \textit{h} = 0.1 and 10000 agents. $x$-axis plots $u_m$ and $y$-axis plots various $u_m$.}
\label{fig5}
\end{center}
\end{figure}

The fig. \ref{fig5} shows the maps for main dimension. 

\begin{figure}[ht]
\begin{center}
\includegraphics[width=10cm]{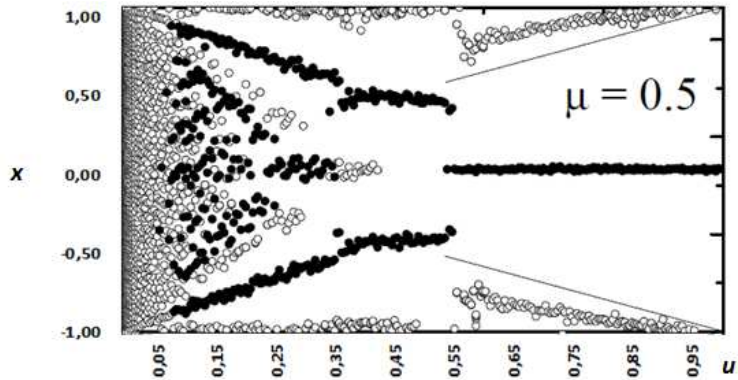}
\caption{Distribution of agent’s opinions \textit{x} (on the $y$-axis) as a function of  attraction threshold \textit{u} in $x$-axis for $\mu$ = 0.5 - extracted from \cite{Laguna2004}, figure 2. White dots indicate minor clusters, whereas black dots indicate major clusters, with a population larger than 1000 agents. The lines enclose the basin of attraction of the state
with a single cluster. The results of a single realization are shown in each panel for each value of $u$ on the $x$-axis.}
\label{fig6}
\end{center}
\end{figure}

For $u_m$ $>$ 0.5 corresponding to only one major cluster, we observed a polarized norm within a particular angle-shape of couples of ($u_m$, $u_s$) values. The trajectory shown in the Figure \ref{fig2} occurs in this area which corresponds to a norm change from moderate to extreme. The white area indicates agents that have adopted a moderate norm and then remain unaffected by extreme minor clusters. The angle-shape is due to the speed of convergence which, depending on the value of uncertainty, leaves the agentsat various distances from the center without being influenced by the other converging agents. This was pointed out by \cite{Laguna2004} in their study of the emergence of minor clusters in a bounded confidence model implemented in an agent-based model. Figure \ref{fig6}, which is extracted from their work, clearly shows that in the area corresponding to one major centered cluster (on the right of the graph), the location of the minor cluster, not so far from the center at the beginning, strays further away as $u$ increases. Thus, when the minor clusters are closer on one dimension to the HSI agents shifting towards them, they can be influenced quicker by HSI agents and join the moderate central cluster before the HSI agents have attracted the majority of non-HSI agents of the moderate central cluster to the extreme. When the minor clusters are further, they almost cannot be influenced by HSI agents before the moderate major cluster becomes more extreme. This indicates that this dynamics is probably sensitive to the proportion of HSI agents in the population.

On the left part of the figure, for $u_m$ $<$ 0.55 corresponding to more than one major cluster, we observed norms that polarized for $u_m$=0.45 and $u_s$ from = 0.3 to 0.75 (see the darkest gray of the left part of the figure). This $u_m$ value indicates that the stable state has two major clusters. The darker area thus corresponds to a bipolarization of the population. For lower and larger values of $u_s$, the population gathered in two moderate clusters is not sensitive to the minor  extreme clusters.

Note that this kind of polarization on the main dimension only occurs when the population is mixed, composed of HSI and non-HSI agents. It does not occur with a simple 2D bounded confidence model, nor with a model like \cite{HuetDeffuant2010} with a population composed only of HSI agents.

\subsection{When does the norm change occur on the secondary dimension?}

\begin{figure}[ht]
\begin{center}
\includegraphics[width=10cm]{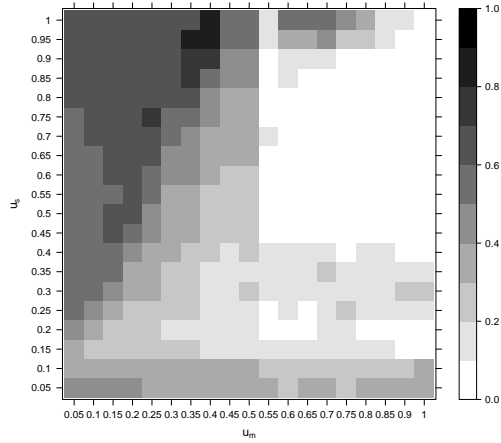}
\caption{Variation of the average absolute opinion on secondary dimension for \textit{h} = 0.1 and 10000 agents. $x$-axis plots $u_m$ and $y$-axis plots various $u_m$.}
\label{fig7}
\end{center}
\end{figure}

Figure \ref{fig7} shows the maps for the secondary dimension. 

For $u_m$ $>$ 0.5 corresponding to only one major cluster, we observed a polarized norm for (a) large $u_s$ $>$ 0.9 and (b) small $u_s$ $<$ 0.15. The trajectory shown in figure \ref{fig4} occurs in this area, for a large $u_s$. This is where we observe a norm change: from moderate to extreme. What occur in this area for small $u_s$ corresponds to the trajectory shown in the figure \ref{fig3}, which is a very long transitory polarization on the secondary dimension. We have not done a long-enough simulation to observe the final moderate consensus, so the centralization is not yet total at the end of our simulations. The white and light-grey areas indicate a total centralization where agents have adopted a moderate norm and then remain unaffected by extreme minor clusters.

\begin{figure}[ht]
\begin{center}
\includegraphics[width=12cm]{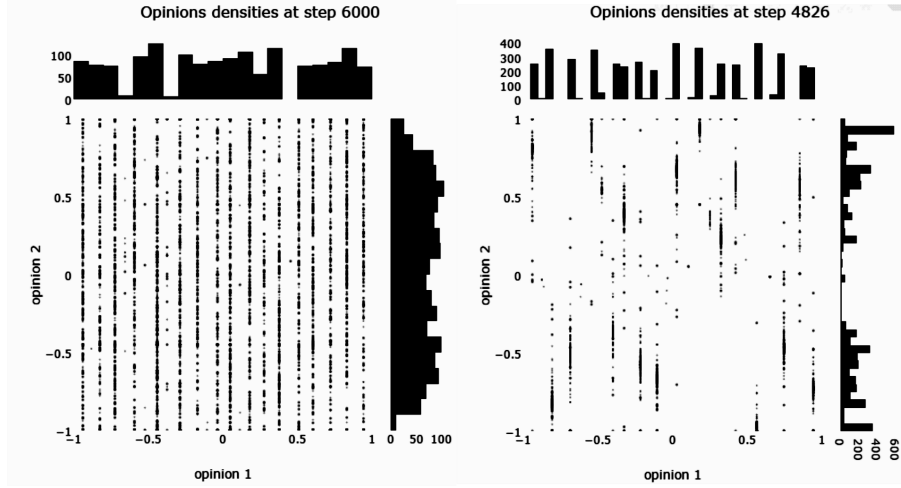}
\caption{Dynamic equilibrium state for 5000 agents, $u_m$ = 0.05 and $u_s$ = 1.1, for \textit{h} = 0.7 on the left and for \textit{h} = 0.05 on the right. A small \textit{h} (on the right) allows agents to cluster on the secondary dimension even if HSI agents continue moving, whereas for a large \textit{h} the agents are always moving and unsatisfied (on the right). This clearly shows that the smaller proportion of HSI agents allows non-HSI agents to cluster despite the bounded property of the opinion space. The HSI agents are located on the extrema of the secondary dimension and in the center. Clusters, mainly composed of non-HSI agents are stable even if close on the main dimension. Since HSI agents, a minority, are attracted by their related clusters of non-HSI agents, they remain close to them on the secondary dimension despite their discomfort with also being close on the main dimension with other clusters and/or on the borders of the secondary dimension. They have pushed their own clusters away from the center of the secondary space, but there are too few of HSI agents to make them move continuously.}
\label{fig8}
\end{center}
\end{figure}

On the left part of the figure, for $u_m$ $<$ 0.55 corresponding to more than one major cluster, we observed a polarization of norms when $u_s$ is large enough compared to $u_m$ (darker areas). Note there is not only a polarization on the left corner (darker areas) but also a centralization (mid-grey area around the diagonal of the left part). The population states in this centralization area is composed of clusters of agents having a similar opinion that is closer to the centered opinion 0 compared to the expected moderate behavior. This expected behavior corresponding to the case of HSI agents having no impact is reached in the transitory area between the darker top triangle and the mid-grey diagonal [see 3.3 for details on how to interpret the average absolute opinion values]. These polarization and centralization dynamics on the secondary dimension, as well as the position of the moderate frontier between the two, can be properly explained. The frontier is defined by $u_s$=2 $u_m$. Indeed, in a pure non-HSI-agent population, cluster localization is exclusively due to the attraction process, as agents gather each other in areas of $u_s$ or $u_m$ widths. So two clusters, located in two different areas, are distant from two $u_s$ or $u_m$ (depending on the dimensions we look at). Conversely, when the population contains some HSI agents, the attraction on the secondary dimension is no longer bounded by $u_s$ since HSI agents are attracted and attract in return every other agent that is close on the main dimension until they all become members of the same cluster on the secondary dimension too. There is thus a strong tendency to gather in the center of the secondary dimension in a first step. However, in a second step, the rejection by HSI agents of members of other clusters defined by the main dimension pushes away the people from different clusters on the secondary dimension. This is a norm change: from moderate at first, to polarized. However, HSI agents  do not need to differ from more than $u_s$ to become stable. This is why clusters have a closer equilibrium position in the opinion space on the secondary dimension (distant from $u_s$) than expected from a more oriented attraction dynamics (distant from 2 $u_s$). As confirmed by Figure \ref{fig8} at right, this is affected by \textit{h}, which partly rules the initial attraction strength against the rejection process.

\begin{figure}[ht]
\begin{center}
\includegraphics[width=12cm]{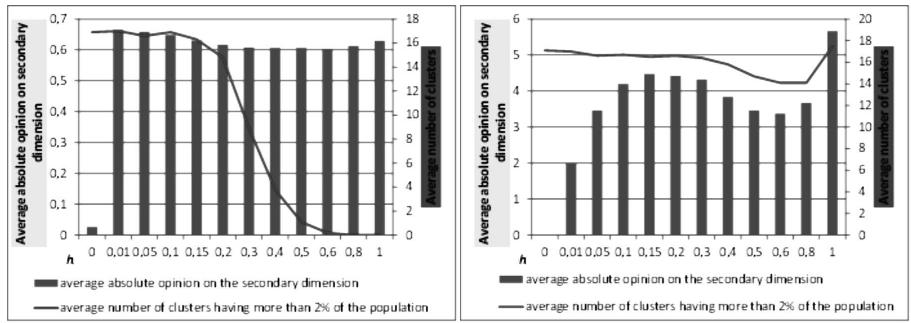}
\caption{Average absolute opinion on the secondary dimension (bars) and average number of clusters containing more than 2\% of the population (line) for various \textit{h} on $x$-axis – on the left for an bounded opinion space and on the right for an unbounded opinion space (5000 agents – $u_m$ = 0.05 and $u_s$ =1.1, 10 replicates). Note that polarization level is the same for all \textit{h} in the bounded space case but sensitive to \textit{h} in an unbounded opinion space. This finding warrants further investigations, which are not in the scope of this paper.}
\label{fig9}
\end{center}
\end{figure}

Overall, for a population containing HSI agents, the frontier between centralization and polarization on the secondary dimension is defined by $u_s$=2$u_m$, regardless of whether the dimension is bounded, because this is the exact value for which the opinion space occupied by the clusters on the main dimension and the second dimension is the same, knowing that the whole space on the main dimension is occupied according to the uniform initial distribution of opinion. For $u_s$ $<$2$u_m$, there is less space occupied on the secondary dimension, which implies a centralization of opinions that are consequently closer to the center than they initially were. For $u_s$ $>$2$u_m$, the clustering process on the secondary dimension requires more space than that occupied on the main dimension. This opinion evolution occurs when the rejection threshold is significantly higher than the attraction threshold. For these values, the rejection process is dominant. The initial attraction on the secondary dimension is weak. Indeed, as the attraction on the main dimension is also weak, many agents stay far from each other. The polarization on this dimension begins before the centralization has really been formed. In this case, people are particularly narrow-minded  about what is important for them, and thus form a lot of groups. Moreover, HSI agents want to be very different from agents of other groups on the secondary dimension. They socially self-define by differentiation to others. For a pure population of HSI agents and for very small $u_m$ values and very large $u_s$ values, agents continue to fluctuate on the secondary dimension without being able to find an opinion allowing them to differentiate from agents of other groups. In the same conditions, for a mixed population with a small enough \textit{h}, agents do not fluctuate. Figure \ref{fig8} illustrates these behaviors and shows (see left) that a mixed population tends to cluster. Indeed, on the left, we observe that for low \textit{h} values, the number of clusters is the same for a bounded opinion space as for an un-bounded opinion space (see on the right of Figure \ref{fig9}). Conversely, on the left of figure \ref{fig9} showing a pure population of HSI agents in a bounded opinion space, for larger \textit{h} values, the number of clusters decreases to zero since the continuous move on the secondary dimension makes the cluster-counting algorithm inefficient.

Note, finally, that the polarizations for $u_m$ $>$ 0.5 on the secondary dimension only occur when the population is mixed, composed of HSI and non-HSI agents, at least for $u_s$ $\leq$ 1. It does not occur with a simple 2D bounded confidence model, nor with a model like \cite{HuetDeffuant2010} with a population composed only of HSI agents. The polarization for $u_m$ $<$ 0.55 and a large enough $u_s$ does not occur with a simple 2D bounded confidence model but does occur with a model like \cite{HuetDeffuant2010}  with a population composed only of HSI agents.

\subsection{What about the two dimensions seen simultaneously?}

Overall, from the comparison between the figure \ref{fig5} and the figure \ref{fig7}, one can notice the two types of polarization, on the main and on the secondary dimension, do not appear simultaneously. However, the centralization area on the secondary dimension, delimited by $u_s\simeq 2u_m$, corresponds to very different behaviors on the main dimension: polarization, centralization or no change. This means that while the cohesion increases on the secondary dimension, all the behaviors are possible on the main dimension.

We would intuitively think that the polarization can be explained by the fact the opinion space is bounded. However, this is not the case, as can be seen in Figure \ref{fig10} below showing results for an unbounded attitude space, where we see the same polarization areas on both maps.

\begin{figure}[ht]
\begin{center}
\includegraphics[width=6cm]{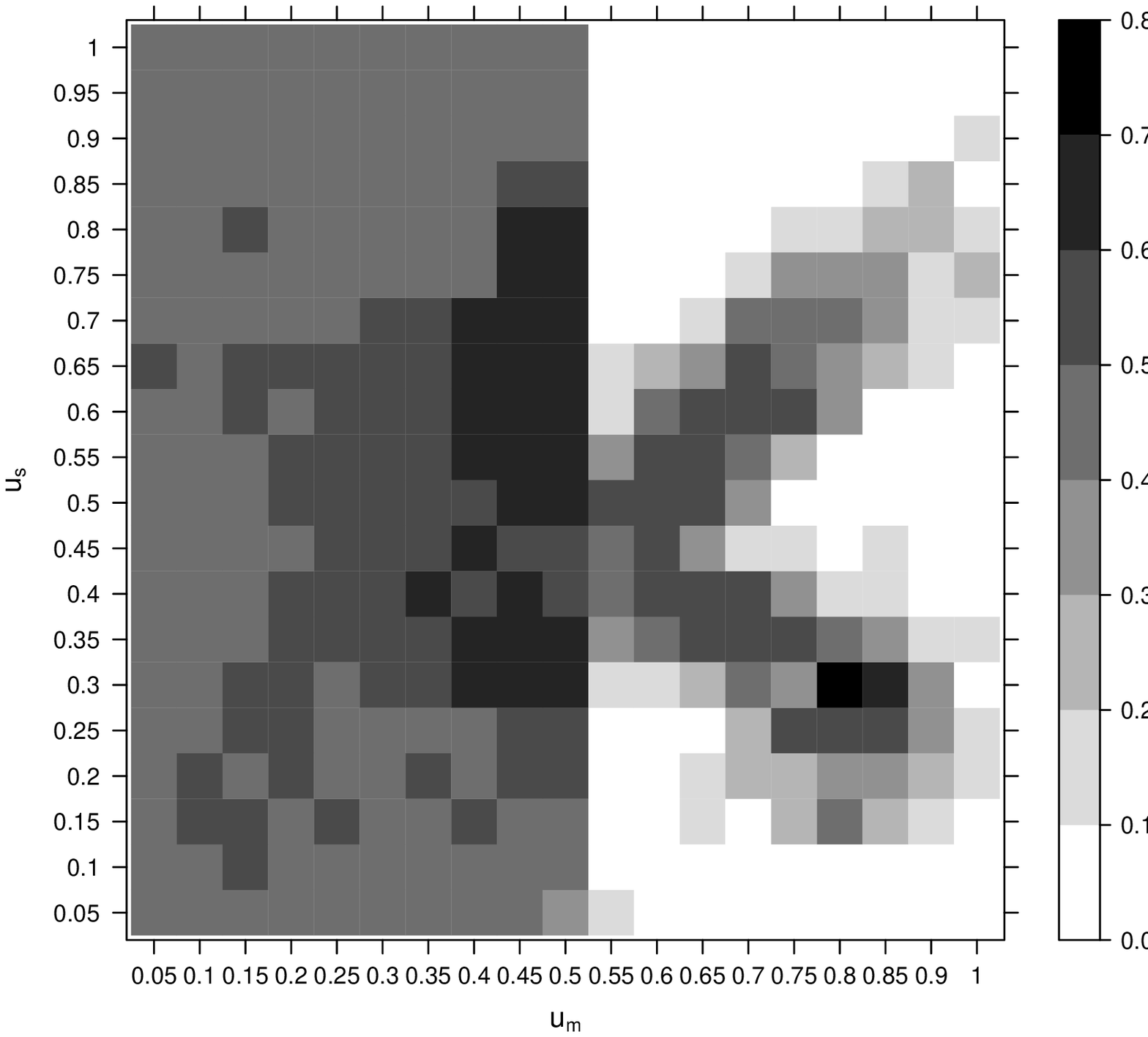}
\includegraphics[width=6cm]{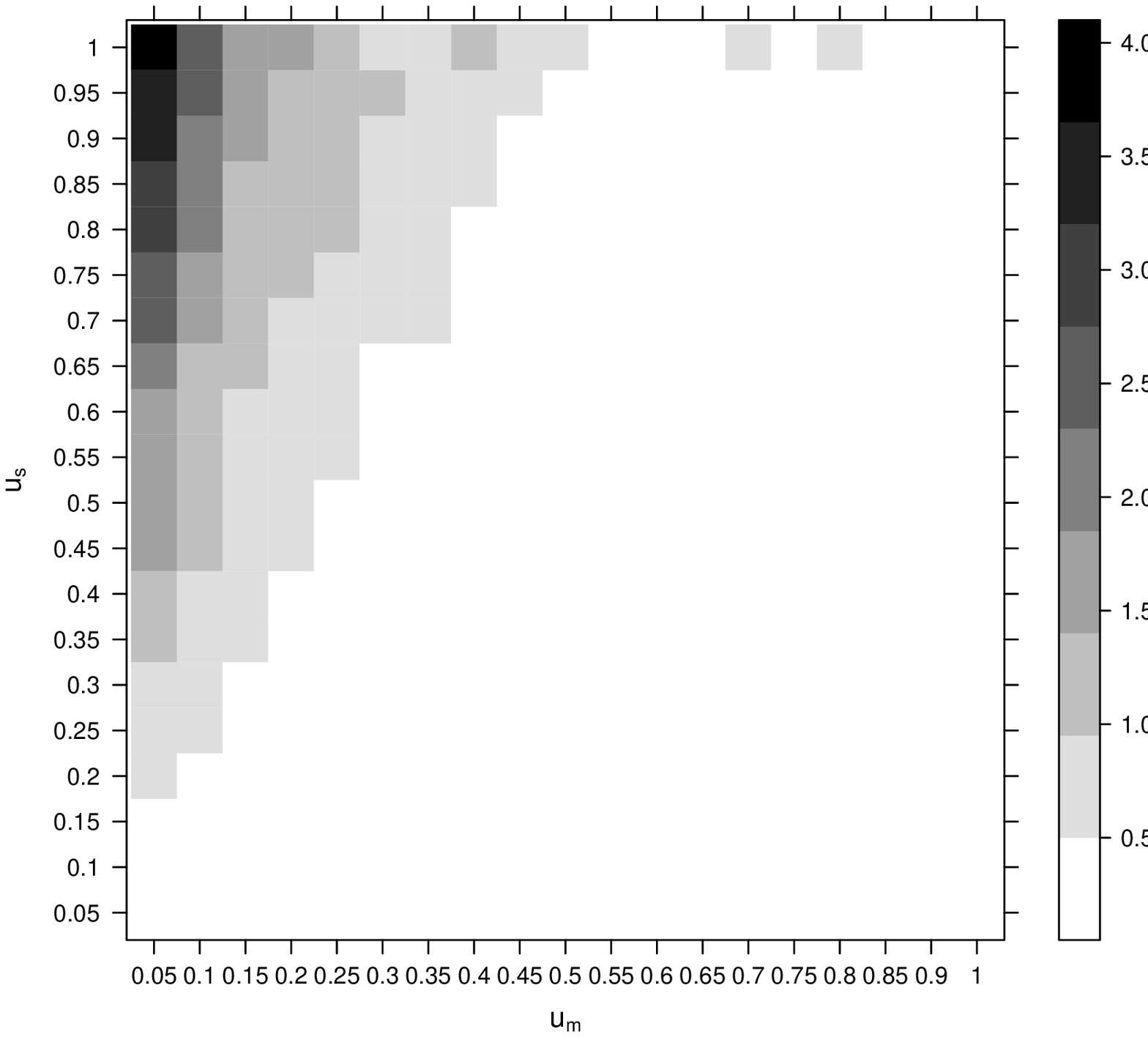}
\caption{Average absolute opinion on the main (left) secondary (right) dimension for 10000 agents evaluating in an unbounded opinion space, $h$=0.1. $x$-axis plots $u_m$ and $y$-axis plots $u_s$. While white (right part on the right figure) or mid-grey (left part of the left figure) charts the area with a moderate polarization [which in fact can correspond to no change in polarization compared to the initialization, but also to a decrease preceding an increase], white on the right part of the left figure) charts the area of total centralization and darker grey charts the area of polarization [which has decreased and then increased, or not changed].}
\label{fig10}
\end{center}
\end{figure}


\section{Conclusions}

Here we study how self-involvement can explain a norm change in an agent-based model. We consider a population composed of two different types of agents: highly-self-involved (HSI) agents having a dynamics based on \cite{HuetDeffuant2010} inspired by a particular set of experiments \cite{WoodPool1996}; non-highly-self-involved agents (non-HSI) with a dynamics based on a 2D bounded confidence model \cite{Deffuant2001}. The dynamics of HSI corresponds to the particular behavior in terms of attitude change of people who are highly self-involved in one dimension and consider it as self-defining. From the modelling point of view, it assumes that one dimension is more stable and more important than the other. This more stable dimension, called main dimension, represents the highly self-relevant issue for the agents, and it rules the attitude change on a secondary dimension. If two agents are close on the main dimension (separated by a distance smaller than $u_m$), then they attract each other on the main dimension and on the secondary dimension, whatever their disagreement on the secondary dimension. If they are far from each other on the main dimension, then proximity on the secondary dimension (at a distance less than $u_s$) is uncomfortable, and generates rejection on this dimension. Proximity on a dimension is defined by comparing the attitude distance with the agent’s threshold on this dimension. For non-HSI agents, neither main nor secondary dimension is more important. Then, to be influenced, the non-HSI agent should consider the peer source as close on the two dimensions simultaneously. One might think this latter condition for influence is hard to meet, yet \cite{WoodPool1996} state that participants who do not consider the source as highly self-relevant do not change their opinions, and experiments reported in \cite{Michinov2003} \cite{Michinov2011} \cite{Montoya2011} \cite{Reid2013} show that the level of attraction for peers depends on the degree of similarity with them.

As expected, simulations show agents gathering in attitudinal groups. The number of groups is defined by how close they are on the main dimension to the rest of the population. Even a few HSI people are enough to minimize the number of clusters.

The moderate cluster(s) built in the central part (i.e. having a moderate norm) of the opinion space at first can later polarize on main issues for some particular couples of values ($u_m$, $u_s$) and adopt a (some) more extreme norm(s). This is due to the interaction of these major moderate cluster(s) with some extreme minor clusters. These extreme minor clusters emerge from the dynamics and are only composed of non-HSI agents. When they appear close enough to major clusters, they get absorbed by them, whereas when they are further in the opinion space, but not far enough for moderate HSI agents to see them as not in-group members, they attract the moderate agents without being influenced by them. Indeed, they are more stable, as they are more demanding on the conditions of closeness to be influenced (i.e. closeness on the two attitudinal dimensions is required). The attracted moderate HSI agents are then attracted by the non-HSI agents located both in the center and in the extremes. However, only the interaction with centered agents is symmetrical. Due to the stochasticity of the model, one of the minor extreme clusters finally has more impact on the moderate agents. Then, at this time, extreme non-HSI agents attract moderate HSI agents who, in turn, will attract the moderate non-HSI agents. This is a long process, since moderate non-HSI agents resist by also attracting HSI agents. However, as extreme non-HSI agents cannot be attracted, they finish by winning, and the previous majority of moderate agents becomes more extreme.

A second type of norm change from moderate to extreme appears on the secondary dimension. It occurs when some agents remain forgotten on the extreme of the main dimension - i.e. non-HSI agents, as well as some HSI agents, located far enough from the center on the main dimension—are rejected on the secondary dimension during the convergence to a moderate opinion of the population majority. These extreme agents, both HSI and non-HSI, form extreme minor clusters on the main attitudinal dimension. The norm change of the clusters is due to HSI agents rejecting other HSI agents from other groups and attracting agents of their own groups. This echoes the theories related to groups and identity, especially self-categorization theory \cite{Mark2003} \cite{FlacheMas2008} \cite{JagerAmblard2004} arguing that agents look for a compromise between difference and similarity to others at group level. This type of polarization depends on rejection. It is more intuitive, since people are more certain about their fundamental issue than their secondary issue.

The appearance of a long transitory in-group polarization associated with large certainty (i.e. small uncertainty) on secondary issues is also of interest. It also perfectly describes how HSI increases the level of in-population cohesion compared to a population of non-HSI. 

\section{Discussion}

Our discussion consists mainly in a comparison to other models of attitude, as recommended in \cite{Flache2017review}. Overall, these first results seem to reconcile different points of view on the possible source of polarization due to minorities: (1) attraction for extreme minority on the main issue, or (2) rejection on the secondary issue.

In contrast with most of the scholarship presented in introduction, the hypotheses of our model do not favor one or the other reason for evolution to polarized norms, but instead integrates both mechanisms and shows that both explanations are compatible and can occur in different contexts.

Indeed, if we focus on the hypothesis that the change of norm is due to an attraction of the majority by a minority, we can compare our work to \cite{Deffuant2006} and \cite{DeffuantWeisbuch2007} and more recently \cite{Mathias2016} or \cite{hegselmann2015opinion} with a slightly different version of the bounded confidence model (these two versions have been properly compared in \cite{Lorenz2007}). In our model, as in these works, the evolution to a polarized norm does not depend on rejection, but only on attraction to an extreme minority on the main issue. The result of our experimental design has been also interpreted with the patterns reported by these authors (see the Appendix). One can draw a parallel with the classical increase of $u$ and proportion of extremists in the population explaining the emerging pattern in the bounded confidence with extremists, and the increase of $u_m$, ruling the number of clusters and $u_s$ ruling how strong the extremists are. The models differ in the hypothesis explaining the strength of the minor extreme cluster: in the bounded confidence model, the strength of the extremist cluster is directly proportional to its size, whereas in our model this relationship is less direct, because the influence requires conditions on both the attitude dimensions. But the main difference is that extremists are endogenously added at initialization in the bounded confidence model with extremists, whereas they emerge in our model. This emergence makes the impact of $u_s$ complex and difficult to disentangle. Note too in Figure 11 in Appendix on the left map that ``bipolarization"' in the darkest gray occurs for lower uncertainty values (i.e. $u_m$) compared to the bounded confidence model with extremists ($u$).

This suggests that an evolution of the fundamental attitudes defining a norm is not due to the rejection mechanism on these attitudes but more to the low self-involvement (i.e. strong uncertainty) of the majority of agents, which makes them easier to influence by the extreme minor groups. In some cases, these minor groups are themselves conversely very resistant to the influence of the moderate majority, which is the condition they need to attract this majority to their extreme attitude. In the bounded confidence model with extremists, this resistance comes from strong certainty, whereas in our model it is due to a higher similarity requirement to trigger influence. In fact, both explanations can be seen as complementary. Both of them are compatible with the attitude strength-related literature which argues that extremity, importance and certainty are key factors in the capacity to resist influence \cite{PettyAndKrosnick1995} \cite{BizerEtAl2004}.

Turning to focus on the second possible explanation for evolution of a norm, i.e. willingness to differentiate, we can also compare our model to the classical patterns of the bounded confidence model with extremists. An evolution of the moderate norm to a polarized norm occurs due to rejection on the secondary dimension in our model. We can see in figure 11 in Appendix on the right map that our model is able to produce a ``single extreme"' and ``bipolarization"', again for lower uncertainty values ($u_m$) than the bounded confidence model with extremists. It also shows a ``polarization"' when there are numerous clusters that, to our knowledge, do not appear in the bounded confidence model with extremists.

Here then, overall, and standing apart from most of the models considering a rejection process presented in \cite{Flache2017review}, we show in a unique model the various behaviors and explanations related to the emergence of extreme clusters and moderate to polarized norms. Moreover, we show that our conclusions are also valid when the attitude space is not bounded. 

However, our results are dependent to a number of assumptions:
\begin{itemize}
	\item as already stated, the emergence of extreme minority clusters is due to the speed of the converging process (i.e. due to the parameter $\mu$); nothing occurs when the speed of convergence to the center is lower ;
	\item the model assumes a uniform initial opinion distribution. Compared to other distributions, depending on factors such as the standard deviation of a normal initial distribution, it yields a lower or higher number of extremists that might exacerbate the emergence and influence of extreme clusters when these clusters are due to the speed of the attraction process;
	\item the interaction network is complete: every agent can try to interact with any other agent. Regardless of whether this assumption is plausible, it can be expected to have a positive effect on the number of interactions before convergence, and perhaps on the number of agents who remain extremists due of the convergence dynamics;
	\item the most important dimension is the same for every agent, so the dimension for unconditional attraction and rejection is the same for everyone, which makes it quite difficult to expect what would occur if the most important dimension varies from one agent to another;
	\item homogeneous $u_m$ and $u_s$ thresholds, which might make opinion clusters more stable than they would be if agents had different thresholds;
\end{itemize}

The next step is to study the sensitivity of our results to these assumptions.


\section*{Acknowledgments}
The authors thank the Auvergne Region in France for providing grant funding (``Emergent themes" 2015, project Associatione), and Guillaume Deffuant (Irstea Lisc) for providing valuable help.


\bibliographystyle{ws-acs}
\bibliography{mybiblio}


\appendix

\section{Results of the experimental design expressed in terms of patterns of the bounded confidence model with extremists}

\begin{figure}[th]
\begin{center}
\includegraphics[width=6cm]{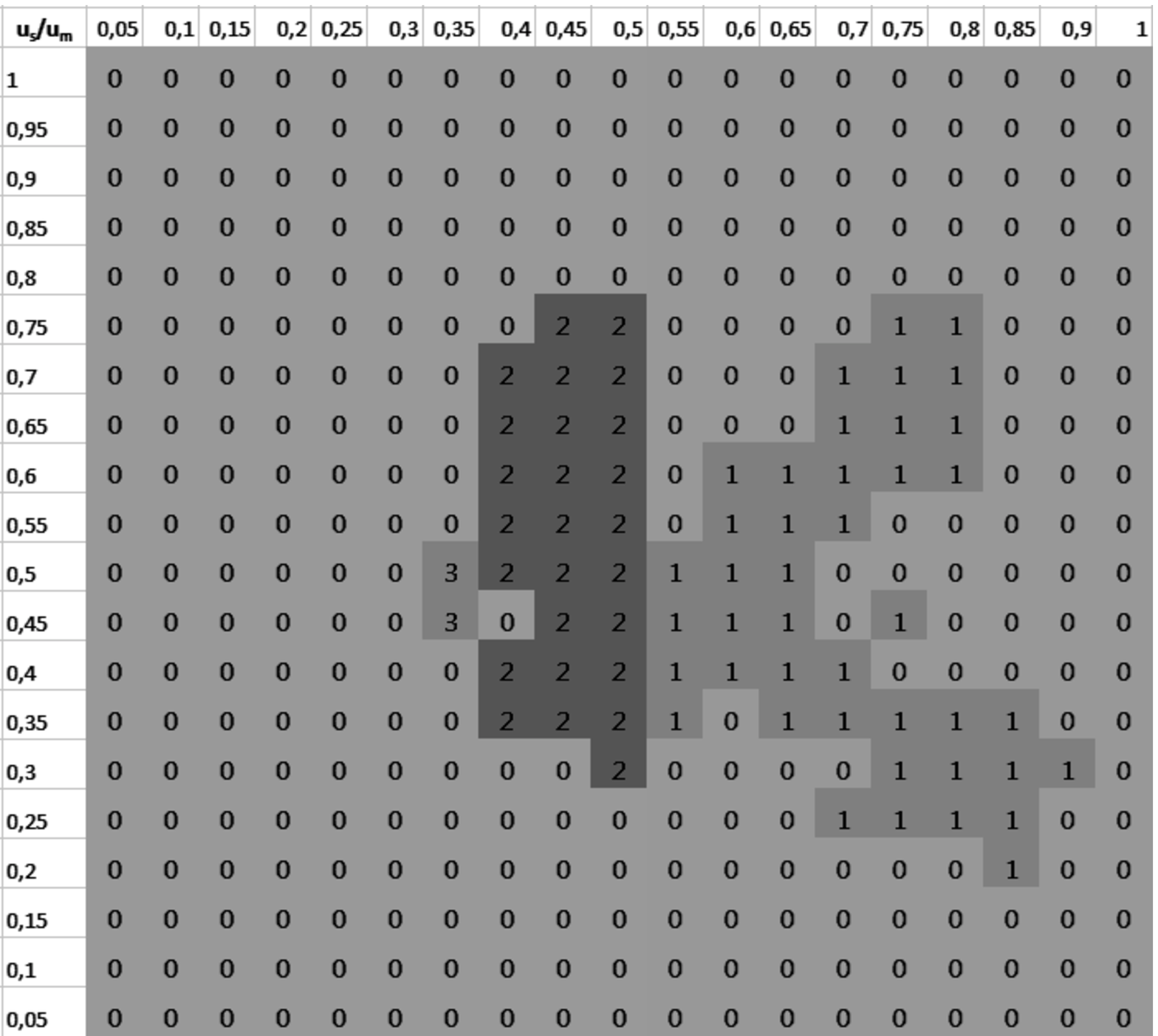}
\includegraphics[width=6cm]{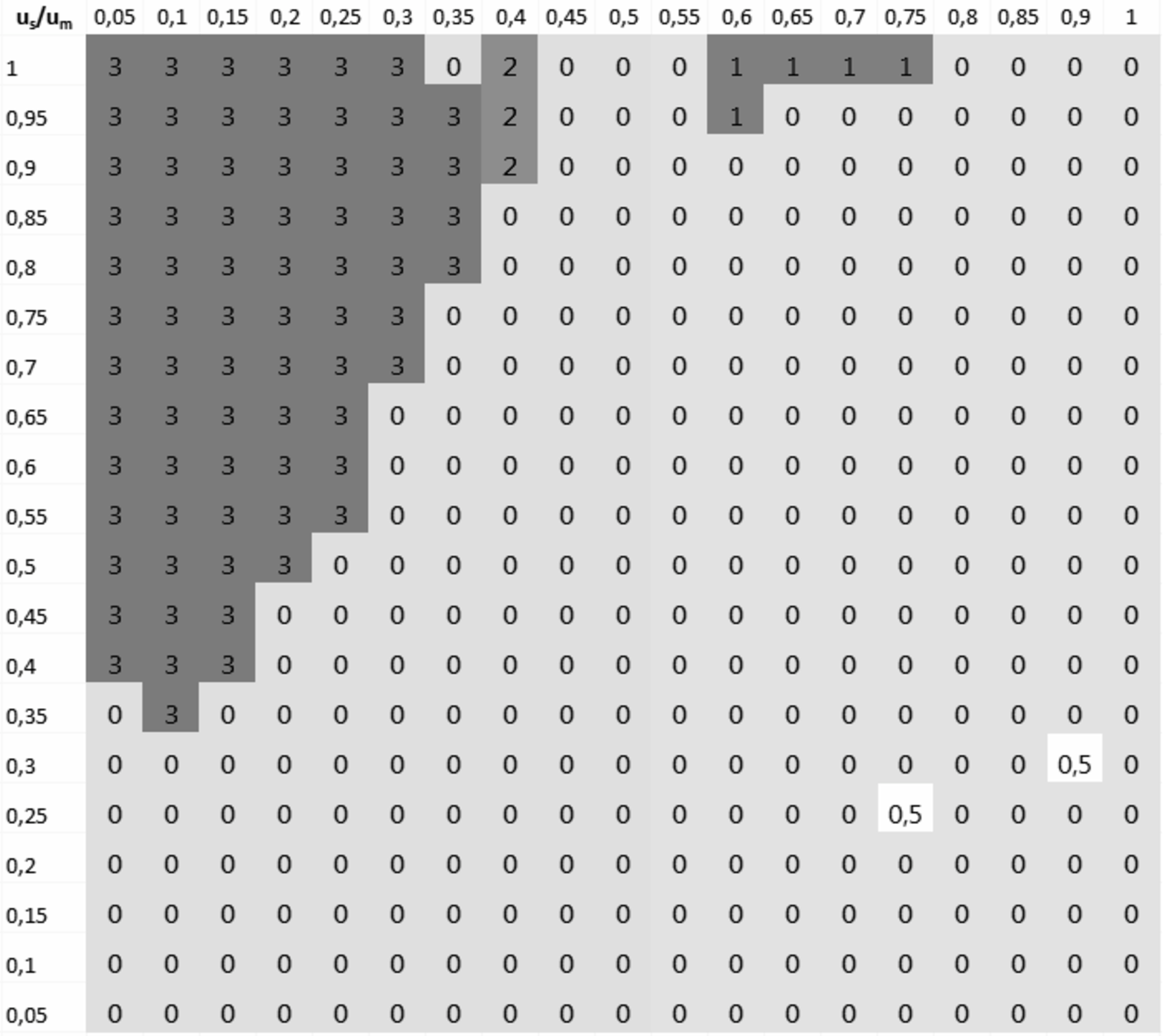}
\caption{Diagnosis on the main dimension (on the left) and on the secondary dimension (on the right), \textit{h} 0.1: 0 (or 0.5) means single moderate cluster; 1 means single extreme cluster; 3 means several polarized clusters; 4 means several moderate clusters; 2 means bi-polarization (two polarized clusters). For each map, $x$-axis plots $u_m$ increasing from right to the left, and $y$-axis plots $u_s$ decreasing from top to the bottom.}
\label{fig11}
\end{center}
\end{figure}

Figure \ref{fig11} presents the map indicating the patterns output from the model for the experimental design presented in section 3.3. A type of convergence, i.e. a pattern, is defined by at least 50\% of replicates leading to this type. The patterns are:

\begin{itemlist}
\item the large majority of agents gather in a single moderate cluster - number 0 on the map;
\item the large majority of agents gather in a single polarized cluster - number 1 on the map; 
\item the agents are gathered in different groups of attitudes or opinions and these groups are located further from each other than expected by the bounded confidence model - number 3 on the map; 
\item bi-polarization, a subcategory of the pattern above, in which the agents are gathered in two major groups of attitudes or opinions and these two groups are located further from each other than expected by the bounded confidence model - number 2 on the map ;
\item the agents are gathered in different groups of attitudes or opinions and these groups are moderate enough since they are located at the same distance from each other as expected by the bounded confidence model - number 4 on the map;
\end{itemlist}
	
The polarization on the main and on the secondary issue does not, at a stable state, seem able to occur simultaneously in most cases. However, a long transitory polarization on the two dimensions is definitely possible for $u_m$ and $u_s$ values in the bottom right corner.

Moreover, the map does not show the state recently pointed out by \cite{Mathias2016} in which non-extremist agents continue to fluctuate without being able to find a stable attitude due to a very large uncertainty and their attraction for the extremists agents. This is because it does not appear with \textit{h} 0.1, but we know from \cite{HuetDeffuant2010} that when \textit{h} is larger, in fact 1 in this paper, this kind of state also appears. However, it differs in the way it can be explained from the dynamics, since it is not only associated to the uncertainty value but also to the rejection process.

\end{document}